\documentstyle[aps,psfig]{revtex}
\newcommand{\grad}{\vec{\nabla}}
\newcommand{\eqref}[1]{(\ref{#1})}
\newcommand{\k}{{\bf k}}
\newcommand{\p}{{\bf p}}
\newcommand{\pf}{{\bf p}_f}
\newcommand{\vf}{{\bf v}_f}
\newcommand{\q}{{\bf q}}
\newcommand{\R}{{\bf R}}
\newcommand{\ber}{\begin{eqnarray}}
\newcommand{\eer}{\end{eqnarray}}
\newcommand{\Tr}{{\bf{Tr}}}
\newcommand{\hg}{{\hat g}}

\begin{document}

\title{The Effect of Surfaces on the Tunneling Density of States of an \\
       Anisotropically Paired Superconductor}
\author{L.J.~Buchholtz}
\address{Department of Physics,
         California State University,
         Chico,~~Chico, CA 95929, USA}
\author{Mario Palumbo, D. Rainer}
\address{Physikalisches Institut,
         Universit\"at Bayreuth,
         D-95440 Bayreuth, Germany}
\author{J.A.~Sauls}
\address{Department of Physics \& Astronomy,
         Northwestern University,
         Evanston, IL 60208, USA }

\maketitle

\begin{abstract}

\centerline{{\bf Abstract}}
{\small\noindent
  We present calculations of the tunneling density of states in an
anisotropically paired superconductor for two different sample geometries:
a semi-infinite system with a single specular wall, and a slab of finite
thickness and infinite lateral extent.
  In both cases we are interested in the effects of surface pair breaking
on the tunneling spectrum.
  We take the stable bulk phase to be of $d_{x^2-y^2}$ symmetry.
  Our calculations are performed within two different band structure
environments: an isotropic  cylindrical Fermi surface with a bulk order
parameter of the form $\Delta\sim k_x^2-k_y^2$, and a nontrivial
tight-binding Fermi surface with the order parameter structure coming from
an anti-ferromagnetic spin-fluctuation model.
  In each case we find additional structures in the energy spectrum coming
from the surface layer.
  These structures are sensitive to the orientation of the surface
with respect to the crystal lattice, and have their origins in the detailed
form of the momentum and spatial dependence of the order parameter.
  By means of tunneling spectroscopy, one can obtain information on both
the anisotropy of the energy gap, $|\Delta(\p)|$, as well as on the phase of
the order parameter, $\Delta(\p) = |\Delta(\p)|e^{i\varphi(\p)}$.

 \medskip\noindent
 {\it To appear in J.~Low Temp.~Phys., Vol.~101, Dec., 1995}}
\end{abstract}


\medskip

\section{Introduction}

  In this paper we study the density of states spectrum of a $d$-wave
superconductor in the vicinity of a specularly reflecting surface.
  The anomalous spectral properties of anisotropic superconductors and
superfluids have been discussed in several publications
\cite{buchholtz81,zhang85,hara86,zhang87,tokuyasu88,buchholtz91,kopnin91,hu94,kashiwaya95,tanaka95,nagato95,matsumoto95}.
  The present work logically follows our examination of the system's order
parameter and surface free energy which was presented in
Ref.~\onlinecite{buchholtz95} (from here on we refer to this paper simply
as [I]).
  Together, these studies are part of an ongoing effort to systematically
examine measurable features which are typical of $d$-wave superconductors.
  We hope, as before, to contribute to the fund of theory necessary for
their definitive experimental identification.

  As discussed in [I], one expects that an anisotropic order parameter will
respond strongly to pair-breaking effects near a wall which is, again, why
we have chosen this setting as our arena of study.
  Moreover, the resulting spectrum of excitation states is quite sensitive
to the details of the entire system and is also accessible to a variety of
experimental probes.
  Accordingly, there are any number of spectral dependencies one could
examine which may well yield significant clues as to the true symmetry of
the order parameter.
  We have chosen three for this paper: 1) surface to lattice orientation,
2) realistic Fermi-surface geometries, and 3) finite size effects.
  The results presented here are not in any way intended as exhaustive
studies of any of these areas.
  Instead, we selected out of a rather large collection of calculations a
condensed subset which we felt was either representative of general behavior
or exceptionally suggestive of potential experimental verification.

  We employ the quasiclassical formulation of superconductivity for the
calculations of this paper and refer the reader to [I] for a more detailed
discussion of these methods.
  The equations of particular importance for the current work are collected
in  section II, and the results are displayed in the two subsections of
section III.
  In section IV we discuss various aspects of the physical significance of our
resulting data and we conclude with a summary of the most prominent points.

\section{Quasiclassical Theory}

\subsection{Basic Equations}

  In [I] we presented the fundamental equations of the quasiclassical
theory in the imaginary-energy (Matsubara) representation.
  These considerations provide a sufficient framework for the study of the
thermodynamic properties of the system.
  A discussion of the spectral properties of the system, however, requires
the real-energy version of the Green function equations presented in [I].
  We therefore must compute the retarded propagator, $\hat{g}^R$, which can
be defined in terms of the Matsubara propagator as follows (see, for
example, Ref.~\onlinecite{serene83}):
\begin{equation}
\hat{g}^R\left(\pf,\R;\epsilon\right) =
  \hat{g}\left(\pf,\R;\epsilon_n\right)
  \left.\right|_{i\epsilon_n\rightarrow\epsilon+i\eta} .
\end{equation}
  The $2\times2$ matrix propagator $\hat{g}^R$ obeys a transport equation
of the same form as in the case of imaginary energies:
\ber\label{trans_equation}
\displaystyle
\left[ \epsilon\hat{\tau}_3 - \hat\Delta\left(\pf,\R\right) \, , \,
       \hat{g}^R\left(\pf,\R;\epsilon\right) \right] +
    i\vf\left(\pf\right) \cdot
    \vec\nabla_{\R}\,\hat{g}^R\left(\pf,\R;\epsilon\right)=0 \, ,
\eer
with the same normalization condition,
\ber\label{norm_cond}
\left[\hat{g}^R\left(\pf,\R;\epsilon\right)\right]^2=-\pi^2\, \hat{1} \, .
\eer
  Here $\vf\left(\pf\right)$ is the momentum dependent Fermi velocity, and
$\hat\Delta\left(\pf,\R\right)$ is the superconducting order parameter
(or pairing self-energy).
  For anisotropic superconductors $\hat\Delta\left(\pf,\R\right)$ exhibits,
in general, a nontrivial dependence on the momentum, $\pf$, and the position,
$\R$, which must be calculated self-consistently from the gap-equation.
  The method for carrying out the self-consistent solution for $\hat\Delta$
was discussed in detail in [I] and will not be repeated here.
  Instead, we simply carry over the results obtained in [I] for the
self-consistent order parameter and use them as input in the solution
of equations~\eqref{trans_equation} and~\eqref{norm_cond} for $\hat{g}^R$.

  The propagator $\hat{g}^R$ carries detailed information on the
quasiparticle excitations.
  In the quasiclassical scheme, a quasiparticle state is characterized by
its excitation energy, $\epsilon$, measured from the Fermi energy, and by
a Fermi surface coordinate, $\pf$, which is the projection of the
quasiparticle momentum onto the Fermi surface.
  The total number of quasiparticle states (per spin) with energy,
$\epsilon$, in the energy interval $d\epsilon$, and Fermi momentum, $\pf$,
in the interval $d\pf$ is given by
\ber
dN\left(\pf,\R;\epsilon\right) = N_f \, n\left(\pf\right)
  \nu\left(\pf,\R;\epsilon\right) d\pf \, d\epsilon ,
\eer
where $N_f$ is the total density of states in the normal state as measured,
for example, by specific heat experiments.
  Band-structure effects lead to the anisotropy factor $n(\pf)$, which is
normalized such that $\oint d\pf \; n(\pf)= 1$.
  Finally, the effects of superconductivity on the density of states are
carried by the dimensionless quantity $\nu(\pf,\R;\epsilon)$,
\ber\label{DOS_factor}
\nu\left(\pf,\R;\epsilon\right) = -\frac{1}{2\pi}\Im
  \left\{ \Tr\left[ \hat\tau_3 \,
                    \hat{g}^R\left(\pf,\R;\epsilon\right) \right] \right\} ,
\eer
which we refer to in the remainder of the text as the ``superconducting
DOS factor''.
  In the normal state, $\nu(\pf,\R;\epsilon)=1$.
  In what follows we focus on the calculation of the quantity
$\nu(\pf,\R;\epsilon)$, and, in particular, how its qualitative features
are affected by the presence of a wall.

  We consider two different sample geometries: a semi-infinite system
occupying the space $x>0$, and a slab of finite thickness, $w$, and
infinite lateral extent oriented perpendicular to the $x$-axis.
  In both cases we consider perfectly reflecting (specular) walls.
  Continuity of the propagator along a classical trajectory
requires\cite{ovchinnikov69,kurkijaervi87}:
\ber \label{boundary_cond}
\hg^R(\p_{f,\mbox{in}},\R_{\mbox{wall}};\epsilon) =
\hg^R(\p_{f,\mbox{out}},\R_{\mbox{wall}};\epsilon) ,
\eer
which yields the appropriate quasiclassical boundary condition for a
specular wall.
  An in-coming trajectory is defined as one in which
$\vf(\p_{f,\mbox{in}})\cdot\hat{n}<0$ while an out-going trajectory requires
that $\vf(\p_{f,\mbox{out}})\cdot\hat{n}>0$ (here $\hat{n}$ is taken to be
the surface normal pointing into the sample).
  The out-going momentum vector, $\p_{f,\mbox{out}}$, is then given in terms
of the in-coming momentum vector, $\p_{f,\mbox{in}}$, by requiring that
the momentum component parallel to the surface be conserved upon reflection,
\ber\label{pf_cond}
  \p^{\|}_{f,\mbox{out}} = \p^{\|}_{f,\mbox{in}} .
\eer
  Note that energy conservation is implied in equation~\eqref{boundary_cond}
since $\epsilon_{\mbox{in}}=\epsilon_{\mbox{out}}$.

  We restrict our attention to the case of simple Fermi surfaces in
which any given ``in-coming'' trajectory is {\it uniquely} connected
to an ``out-going'' trajectory via equation~\eqref{pf_cond}.
  This boundary condition would need to be modified for systems with more
complicated Fermi surfaces (e.g.~multiple Fermi sheets, etc.), and
additional phenomenological parameters would be required to determine the
probability of a given in-coming trajectory being reflected into any one of
the multiple out-going channels.

\subsection{Nontrivial Band-structure}

  Before the self-consistent order parameter can be calculated and a
complete solution of equation~\eqref{trans_equation} can be carried out,
we must first specify the form of the Fermi surface, the corresponding
Fermi velocity, $\vf(\pf)$, the anisotropy factor, $n(\pf)$, and the
quasiparticle pairing interaction, $V\left(\pf,\pf'\right)$ (see
equation~(4) in [I]).
  We approach this problem in two different ways.
  First we consider an isotropic cylindrical Fermi surface with a pairing
interaction that can be parameterized in terms of simple trigonometric
functions (as in [I]).
  The second approach, which we carry over from Refs.~\onlinecite{radtke92}
and~\onlinecite{radtke94b}, is to use actual tight-binding fits to the
band-structure of high-$T_c$ materials, along with a pairing interaction
derived from a microscopic spin-fluctuation theory.
  The first of these two approaches was discussed in detail in [I], and we
therefore focus on the second approach in what follows.

  Following Radtke and Norman\cite{radtke94b}, we approximate the
band-structure of the
high-$T_c$ cuprate superconductors by a 2d tight-binding model of the form:
\ber\label{bandstructure}
E(\p) = t_0
     &+& \frac{t_1}{2}\left[\cos{p_x} + \cos{p_y}\right]
      +  t_2\cos{p_x} \cos{p_y}
      +  \frac{t_3}{2}\left[\cos{2p_x} + \cos{2p_y}\right] \nonumber \\
     &+& \frac{t_4}{2}\left[\cos{2p_x} \cos{p_y} + \cos{2p_y} \cos{p_x}\right]
      +  t_5\cos{2p_x} \cos{2p_y} .
\eer
  Given a Fermi energy, $E_f$, we use the relations $E_f=E(\pf)$ and
$\vf=\grad_\p E(\p)|_{E_f}$ to extract the Fermi wavevector, $\pf$, at
every point on the Fermi surface, along with the corresponding Fermi
velocity, $\vf(\pf)$.
  These quantities then allow us to compute the anisotropy factor, $n(\pf)$,
in the usual way.
  Once the band structure data is known, the pairing interaction kernel
can be evaluated using the weak-coupling version of the spin-fluctuation
model of Radtke {\it et.~al.}\cite{radtke92} as follows:
\ber\label{spin_fluc}
V\left(\pf,\pf'\right) = -V^*_0 F\left(\pf-\pf'\right) ,
\eer
where
\ber
F\left(\q\right) = \left[\frac{1}
                         {1 + J_0(\cos{q_xa} + \cos{q_ya})}\right]^2 .
\eer
  One obtains this weak-coupling version of the interaction presented in
Ref.~\onlinecite{radtke92} by integrating-out the high-energy degrees of
freedom\cite{serene83}.
  In doing so one introduces a cutoff, $\epsilon_{co}$, and a renormalized
interaction parameter, $V^*_0$, which differs, in general, from the bare
interaction parameter of Ref.~\onlinecite{radtke92}.
  The renormalized interaction strength, together with the energy cut-off,
can always be eliminated in favor of the superconducting transition
temperature, $T_c$.
  Hence our pairing interaction is determined by two phenomenological
parameters: the transition temperature, $T_c$, and the exchange coupling
parameter, $J_0$.
  The transition temperature is written in terms of the energy cut-off (in
the weak-coupling theory) via:
\ber
T_{c\mu} = 1.13 \epsilon_{co} e^{-1/\lambda_\mu}
\eer
where the $\lambda_\mu$ are dimensionless coupling constants in the various
symmetry channels (see below).

  Once the quasiparticle pairing interaction is known, the superconducting
order parameter can be computed by self-consistently solving the
weak-coupling gap equation (equation~(4) in [I]) in the Matsubara
representation.
  This equation can be recast into a more convenient form by introducing
a complete set of eigenvalues and eigenfunctions of the interaction kernel:
\ber\label{eig_equation}
\lambda_\mu \eta_\mu\left(\pf\right) =
  \oint d\pf' \; n(\pf') \, V\left(\pf,\pf'\right) \eta_\mu\left(\pf'\right) ,
\eer
where the eigenfunctions, $\eta_\mu\left(\pf\right)$, satisfy the
orthonormality condition:
\ber
\oint d\pf \; n(\pf) \, \eta_\mu\left(\pf\right) \eta^*_\nu\left(\pf\right) =
  \delta_{\mu\nu} .
\eer
  If we now express the superconducting order parameter as a linear
combination of these eigenfunctions,
\ber
\hat\Delta\left(\pf,\R\right) = \sum_\mu
  \hat\Delta_\mu\left(\R\right) \eta_\mu\left(\pf\right) ,
\eer
then the gap equation may be rewritten as a series of equations for the
individual gap amplitudes as follows:
\ber\label{gap_equation2}
\hat\Delta_\mu\left(\R\right) = \lambda_\mu
  T\sum_{\epsilon_n}^{\epsilon_{co}} \oint d\pf \; n(\pf) \,
  \eta^*_\mu\left(\pf\right) \hat{f}\left(\pf,\R;\epsilon_n\right) .
\eer
  The eigenfunction, $\eta_\mu\left(\pf\right)$, associated with the
largest (attractive) eigenvalue $\lambda_\mu$ determines the symmetry of
the bulk stable phase, at least near $T_c$.
  We assume that this symmetry persists to all temperatures.
  The effects of the other symmetry components however (whether they are
associated with attractive or repulsive eigenvalues), may become noticeable
in the vicinity of surfaces~\cite{buchholtz95}, defects, in the cores of
vortices, etc.

\begin{figure}
\centerline{\psfig{figure=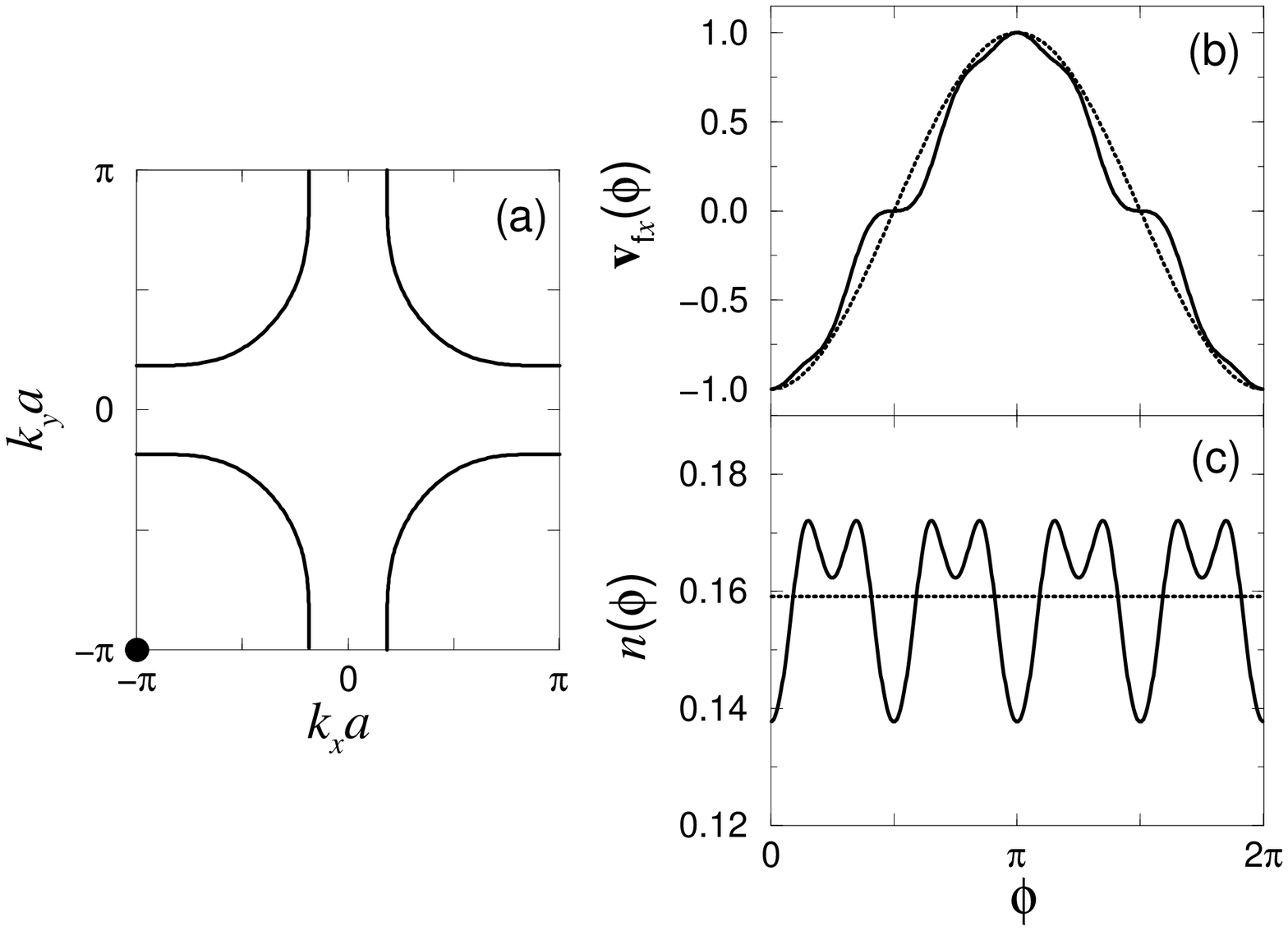,height=3.6in}}
\vspace*{-0.25in}
\begin{quote}
\small
Fig.~1.~~
  Band-structure data: (a) The Fermi surface in the first Brillouin zone.
  We parameterize this surface by an angle $\phi=[0,2\pi]$ taking the
point indicated by the solid dot as the origin.
  In the repeated zone scheme our Fermi surface is then a single continuous
line, around the origin, which is single-valued in $\phi$.
  (b) The $x$-component of the Fermi velocity normalized so that the
maximum is 1.
  (c) The anisotropy factor of the normal state density of states.
  The dotted lines show the corresponding values for the isotropic
cylindrical Fermi surface.
\end{quote}
\end{figure}

  In practice, equation~\eqref{gap_equation2} must be  solved numerically,
which first requires the  numerical evaluation of the eigenvalues,
$\lambda_\mu$, and and eigenfunctions, $\eta_\mu(\pf)$, as defined in
equation~\eqref{eig_equation}.
  Since we would like to expand our order parameter in terms of
eigenfunctions of well defined symmetry, we first project out the pieces of
the interaction matrix, $V\left(\pf,\pf'\right)$, which transform like each
of the four one dimensional representations of the $D_{4h}$ (tetragonal)
group (i.e.~$V_{A_1}\left(\pf,\pf'\right)$, $V_{A_2}\left(\pf,\pf'\right)$,
$V_{B_1}\left(\pf,\pf'\right)$, $V_{B_2}\left(\pf,\pf'\right)$).
  We then use each of these matrices in place of the $V\left(\pf,\pf'\right)$
in equation~\eqref{eig_equation} to solve for the eigenvalues and
eigenfunctions corresponding to each symmetry classification.
  These eigenfunctions then allow us to construct a gap consisting of
specific symmetry components, while the eigenvalues give the coupling
strengths in the various symmetry channels.

  We calculate the Fermi surface data using the parameter values reported
in Ref.~\onlinecite{radtke92} for the YBCO(2,bf) sample (i.e.~$t_0=0.0,
t_1=-1.0, t_2=0.38, t_3=-0.09, t_4=0.0, t_5=0.0$, all in units of $eV$).
  The Van Hove singularity for this system lies at about
$E=-.46eV$\cite{norman95privat}, and we take a Fermi energy slightly above
this value at $E_f=-.30eV$.
  The resulting form of the Fermi surface in momentum space is shown in
Fig.~1a.
  Repeating the Brillouin zone (repeated zone scheme) allows us to
parameterize our Fermi surface by a polar angle $\phi$ around the
$\k=(-\pi/2,-\pi/2)$ point in the Brillouin zone (denoted by a solid black
dot in Fig.~1a).
  In Figs.~1b and~1c we plot the Fermi velocity, $\vf(\phi)$, and the
normalized density of states anisotropy factor, $n(\phi)$, as functions
of $\phi$.
  The factor $n(\phi)$ is obtained from equation (4), yielding
\begin{equation}
N_f \, n(\phi) \, d\phi =
   \frac{1}{(2\pi)^2}\frac{[\pf(\phi)]^2}{|\pf(\phi)\cdot\vf(\phi)|} \, d\phi,
\end{equation}
and is normalized such that $\int_0^{2\pi}d\phi\,n(\phi)=1$.

\bigskip
\begin{figure}
\centerline{\psfig{figure=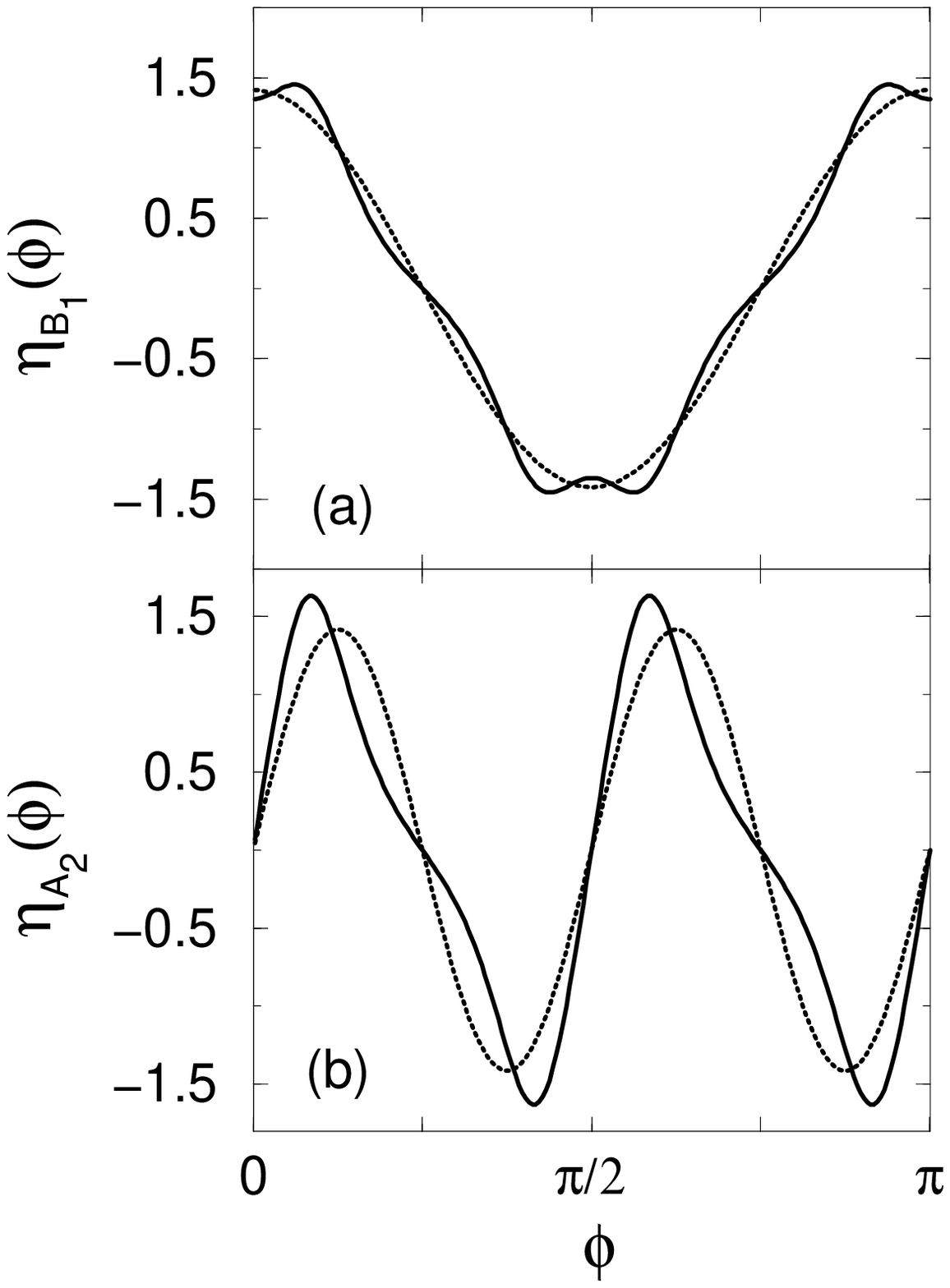,height=3.5in}}
\begin{quote}
\small
Fig.~2.~~
  The momentum space dependence of the $B_1$ and $A_2$ order parameter
components as computed from the weak-coupling spin-fluctuation model
discussed in section~II-B.
  The dotted lines show the corresponding results when a cylindrical
Fermi surface is used.
\end{quote}
\end{figure}

  Once the band-structure data is known, we may then proceed to compute the
spin-fluctuation pairing interaction via equation~\eqref{spin_fluc}.
  Choosing the dimensionless parameters $V^*_0=-1.0$ and $J_0=0.4$, we find
two attractive symmetry channels: $B_1$ and $A_2$.
  The $B_1$ channel has the highest transition temperature while the
$A_2$ channel has a $T_c$ that is reduced by about $3/7$ (the transition
temperatures for the repulsive symmetry channels are set to zero in what
follows).
  Within the dominant symmetry channel, the next to leading coupling
constant is more than two orders of magnitude smaller than the leading
coupling constant and therefore only the leading contribution is considered.
  The angular distribution of the order parameter components corresponding
to the $B_1$ and $A_2$ symmetry channels are depicted in Figs.~2a and~2b.

\section{Results}

  We consider in both of the following subsections a surface geometry of
the same form as in [I].
  The normal to the surface is taken to lie along the $\hat{x}$-axis, which
need not, in general, be parallel to the crystal lattice basis vector
$\hat{a}$, from which we measure all our angles.
  We denote the angle between $\hat{a}$ and the surface normal, $\hat{x}$,
by $\phi_o$, and the angle between $\hat{a}$ and the momentum of an
out-going trajectory, $\p_{f,\mbox{out}}$, by $\phi$.
  This geometry is represented graphically in Fig.~1 of [I].
  We denote the transition temperature for the bulk symmetry channel ($B_1$)
by $T_c$ while the transition temperatures for the other symmetry channels
are denoted explicitly by $T_{cA_1}$, $T_{cA_2}$, $T_{cB_2}$.
  All length scales are measured in units of the coherence length which we
define in the usual way by $\xi=\hbar|\vf|_{max}/\pi\Delta_{max-bulk}$.

\subsection{Semi-Infinite System}

  We consider a semi-infinite system occupying the space $x>0$ with a
perfectly reflecting (specular) wall located at $x=0$.
  The order parameter and free energy of this system were calculated in
detail in [I] (considering an ideal cylindrical Fermi surface)
for several different symmetry parameterizations of the order parameter.
  From our calculations of these quantities in the case of the tight-binding
band-structure and spin-fluctuation pairing interaction discussed in
section~II-B, we conclude that the thermodynamic properties of this system
are quite insensitive to the details of the Fermi surface geometry, as long
as the band-structure does not become too pathological.
  Hence, we do not discuss the stability of the self-consistent order
parameter calculations here.
  Instead, we use these results to calculate $\hat{g}^R$ from
equation~\eqref{trans_equation}, and then focus our attention on the
excitation spectrum.
  This includes the response of the excitation spectrum to an admixture of
subdominant components and to nontrivial band-structure effects.

\bigskip
\begin{figure}
\centerline{\psfig{figure=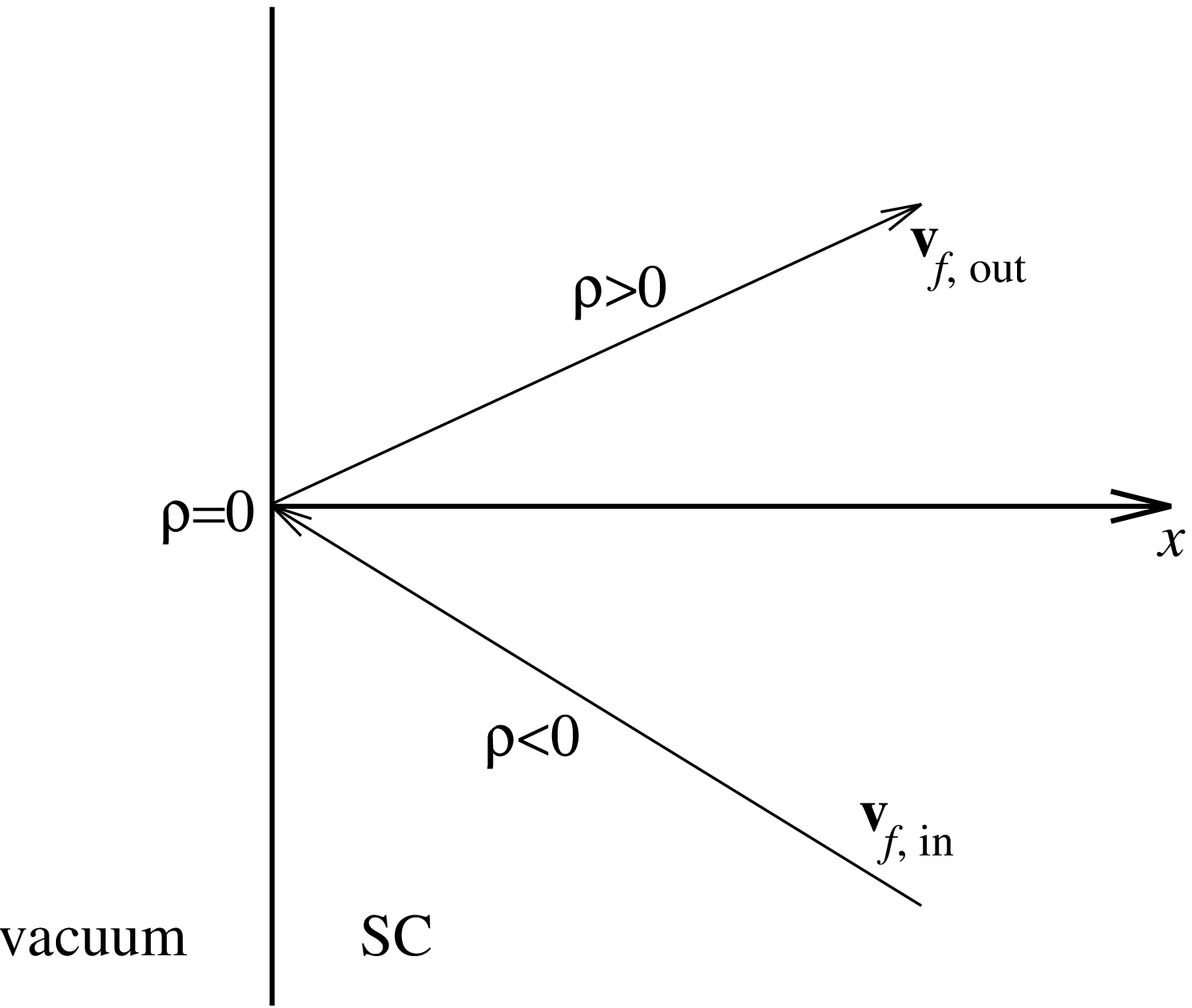,height=2.6in}}
\begin{quote}
\small
Fig.~3.~~
  A typical quasiparticle trajectory reflecting from a specular wall in
the tight-binding model discussed in section~II-B.
  The in-coming leg of the trajectory is indicated by a negative value
for the trajectory parameter, $\rho$, with the point of reflection being
at $\rho=0$.
  Note that for an anisotropic Fermi surface the out-going angle is not,
in general, equal to the in-coming angle.
\end{quote}
\end{figure}

  In Fig.~4 we show some representative curves for the
superconducting DOS factor (defined in equation~\eqref{DOS_factor}) at
$x=0$, together with profiles of the order parameter along the corresponding
``trajectories'' (i.e.~the trajectory of a classical particle approaching
the surface with velocity $\vf(\p_{f,\mbox{in}})$ and leaving with velocity
$\vf(\p_{f,\mbox{out}})$).
  These curves were calculated for the case of a cylindrical Fermi surface
with an order parameter of the form $\Delta\sim k_x^2-k_y^2$.
  A negative value for the trajectory parameter $\rho$ corresponds to the
in-coming segment of the trajectory, while the point $\rho=0$ corresponds
to the point of reflection (a typical trajectory is depicted in
Fig.~3).
  Note that a characteristic ``edge'' appears in the superconducting DOS
factor corresponding to the {\it magnitudes} of the asymptotic values of
the order parameter at $\rho=\pm\infty$.
  For the $\phi=0\deg$ and $\phi=45\deg$ trajectories, these values coincide
so that we find only a single feature in each case at
$\epsilon=(1/2)\Delta_{max-bulk}$ and at
$\epsilon=(\sqrt{3}/2)\Delta_{max-bulk}$ respectively.
  Both the $\phi=20\deg$ and the $\phi=45\deg$ trajectories exhibit a single
pole (or bound state) at $\epsilon=0$.
  These states were discussed by Hu\cite{hu94}, and are a consequence of the
Atiyah-Singer index theorem\cite{atiyah75}.
  It should be emphasized that the presence of these states is a
robust feature that will {\it always} occur ({\it at} $\epsilon=0$) whenever
the local order parameter (which is real in our gauge) changes sign
asymptotically along a trajectory.
  The additional pole found just below the continuum in the $\phi=0\deg$
curve, on the other hand, can be interpreted as a bound state in the
``potential well'' formed by the order parameter along the $\phi=0\deg$
trajectory.
  This is not a robust feature, and will only occur if the well has
sufficient depth and sufficient width.
  The location and weight of this feature therefore contain information
on both the magnitude and the width of the order parameter suppression near
the surface, which are, in turn, directly related to the degree of anisotropy
in the gap.

\bigskip
\begin{figure}
\centerline{\psfig{figure=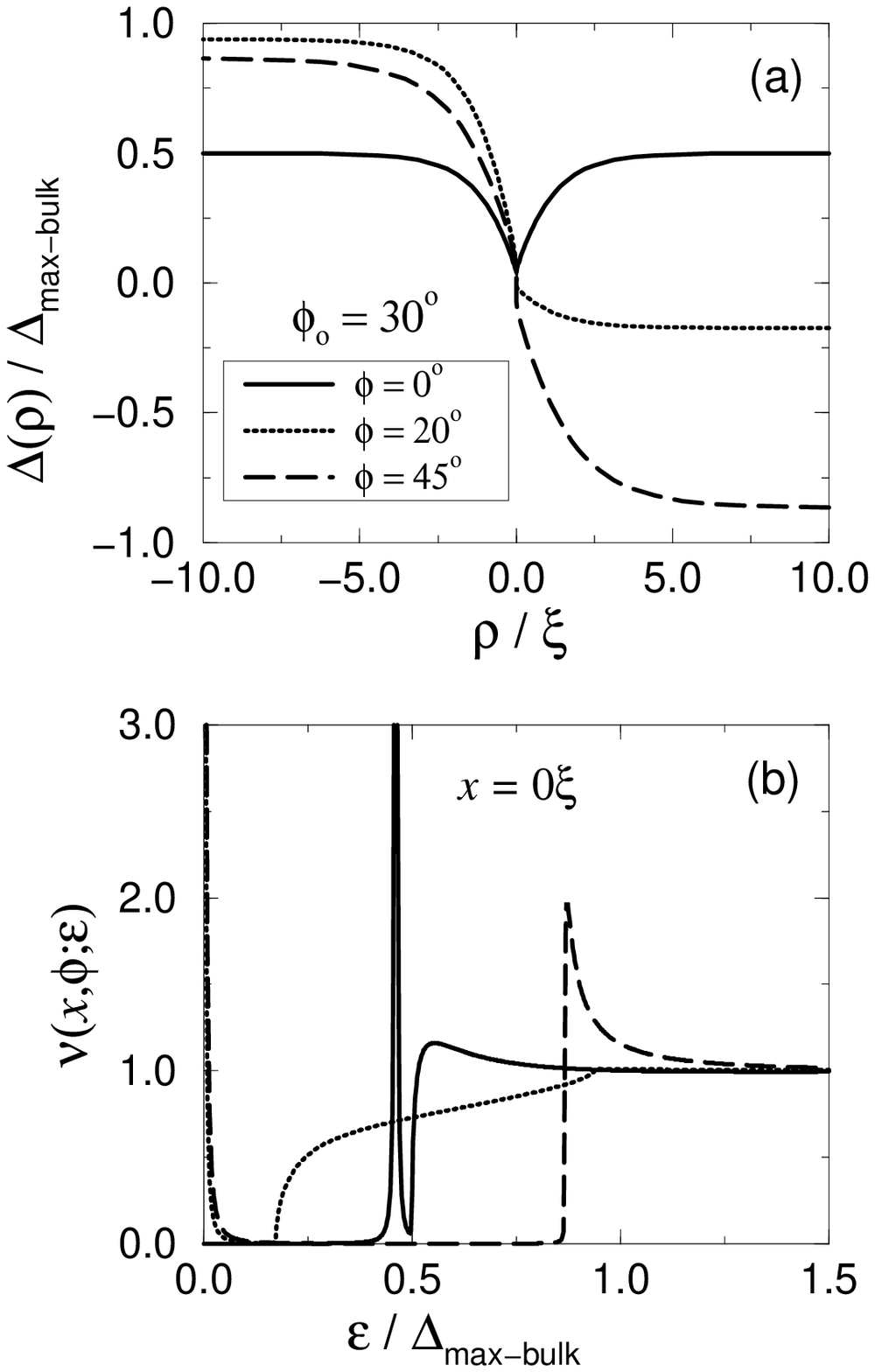,height=3.6in}}
\begin{quote}
\small
Fig.~4.~~
  Order parameter and density of states in a semi-infinite system
(cylindrical Fermi surface, $B_1$-symmetry): (a) the local order
parameter for various (out-going) trajectory angles, $\phi$, and
(b) the superconducting density of states for the same trajectory angles.
  The lattice to surface orientation is taken to be $\phi_o=30\deg$, and
the temperature is $T=0.4T_c$.
\end{quote}
\end{figure}

\bigskip
\begin{figure}
\centerline{\psfig{figure=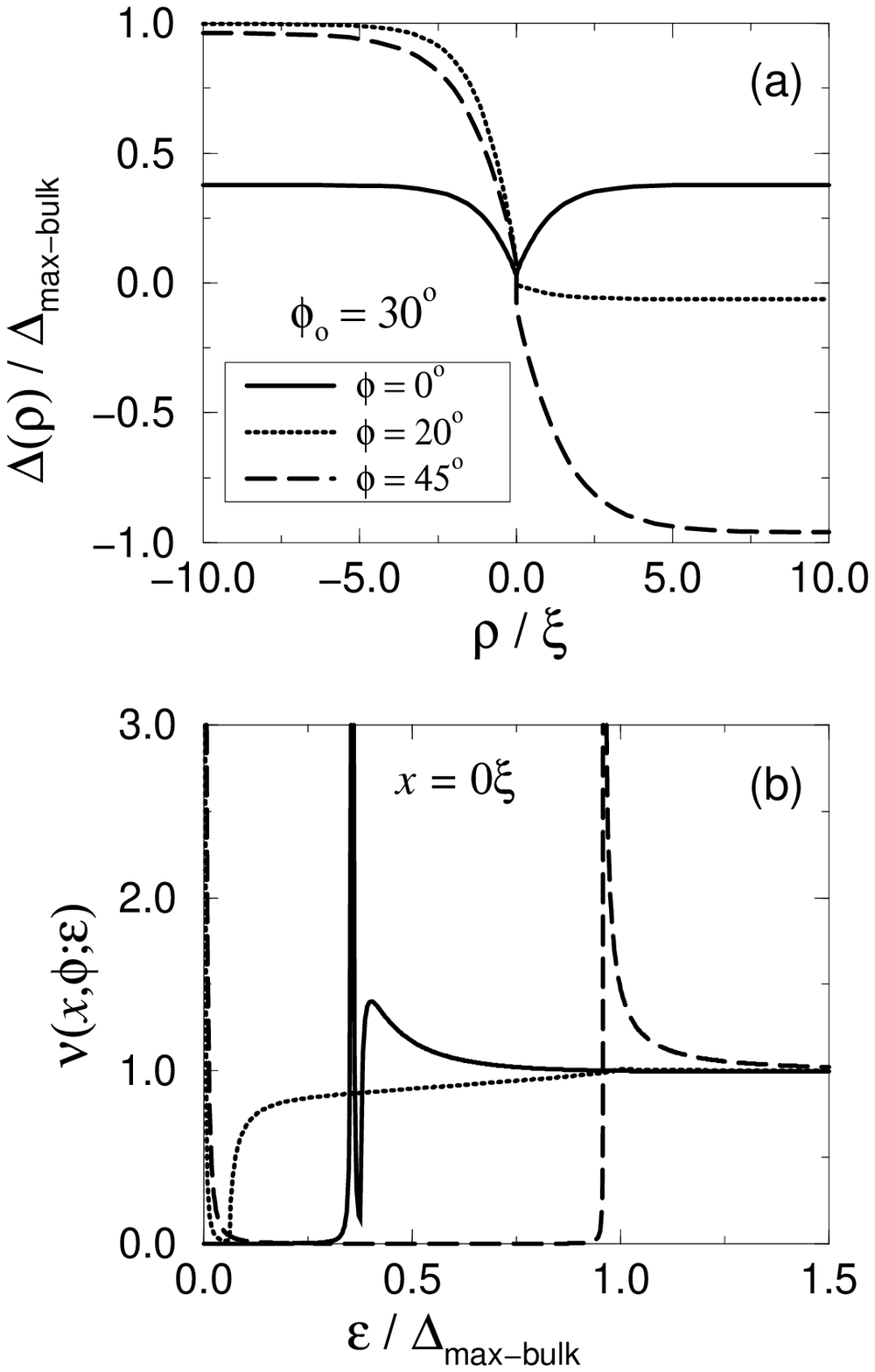,height=3.6in}}
\begin{quote}
\small
Fig.~5.~~
  Order parameter and density of states in a semi-infinite system
(tight-binding Fermi surface, $B_1$-symmetry): (a) the local order
parameter for various (out-going) trajectory angles, $\phi$, and
(b) the superconducting density of states for the same trajectory angles.
  The lattice to surface orientation is taken to be $\phi_o=30\deg$, and
the temperature is $T=0.4T_c$.
\end{quote}
\end{figure}

  For essentially all surface to lattice orientations, the inclusion of
nontrivial band-structure data results in rather minor effects on the
density of states.
  In Fig.~5 we show the superconducting DOS factor
for the same parameter set as in Fig.~4, only now we
consider the band-structure data and pairing interaction discussed in
section~II-B.
  We have suppressed the presence of the attractive $A_2$ symmetry channel
by artificially setting its transition temperature to zero; thus
Fig.~5 corresponds to a $B_1$-symmetry order parameter
with a momentum dependence of the form given in Fig.~2a.
  Other than a redistribution of the location of the gap features,
Fig.~5 is remarkably similar to Fig.~4.
  The fact that the gap features reorganize in this case is understandable
from Fig.~2a where we see that the magnitude of the order
parameter for most momentum angles has been slightly altered.
  However, the key point is that for a single-band Fermi surface of the
form discussed in section~II-B, we expect fairly little alteration in the
results from the simple cylindrical Fermi surface model.
  Such Fermi surfaces have recently been obtained in connection with many of
the high-$T_c$ systems through the use of 2-dimensional tight-binding
models\cite{radtke94b}, and therefore may be considered as reasonable models
for the band-structure in some of the cuprate systems.

\bigskip
\begin{figure}
\centerline{\psfig{figure=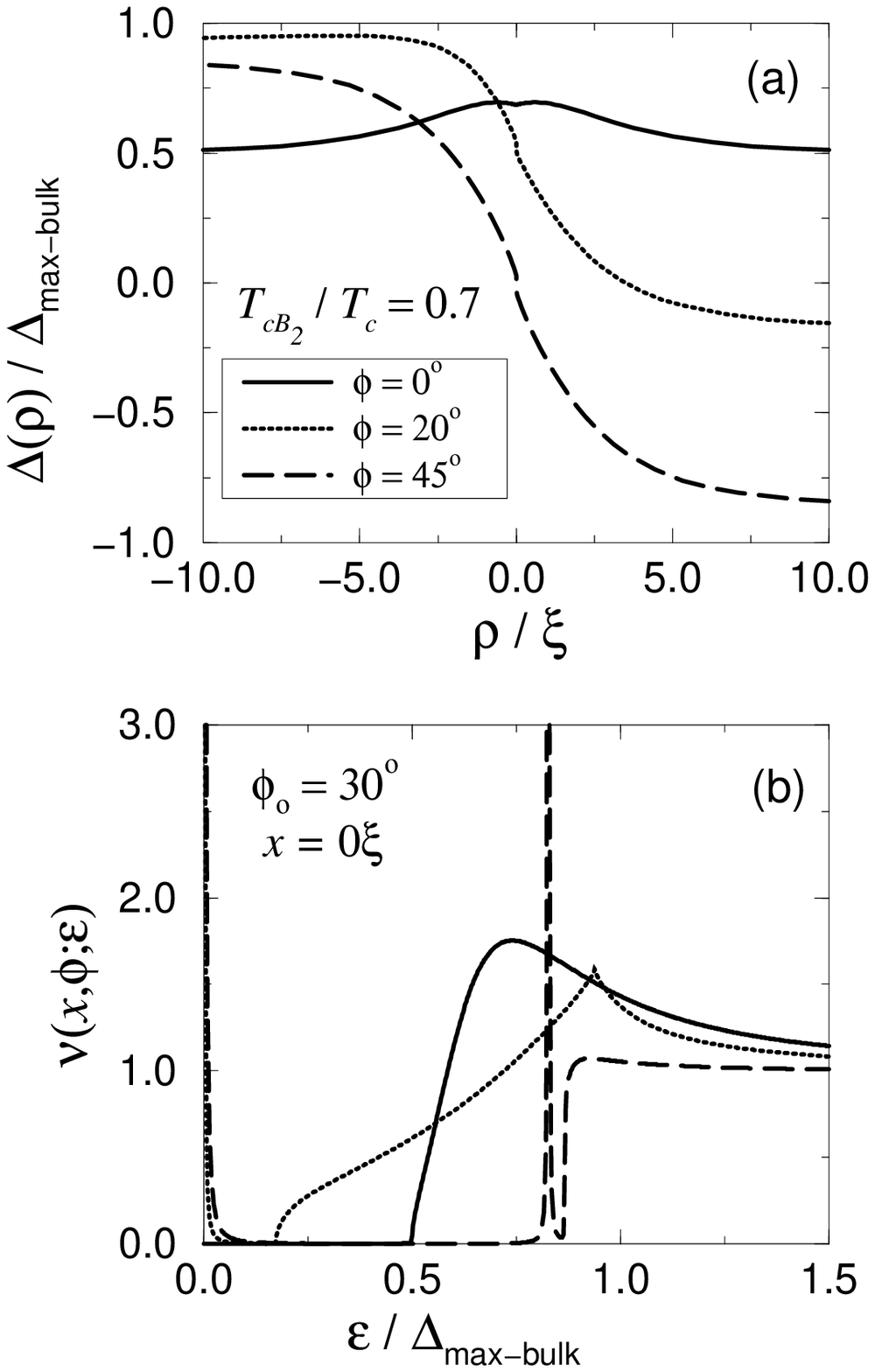,height=3.6in}}
\begin{quote}
\small
Fig.~6.~~
  Order parameter and density of states in a semi-infinite system
(cylindrical Fermi surface, $B_1+B_2$-symmetry): (a) the local order
parameter for various (out-going) trajectory angles, $\phi$, and
(b) the superconducting density of states for the same trajectory angles.
  The lattice to surface orientation is taken to be $\phi_o=30\deg$, and
the temperature is $T=0.4T_c$.
\end{quote}
\end{figure}

  We have also made an extensive study on the effect of the mixing of
sub-dominant symmetry components with the stable bulk ($B_1$-symmetry) phase.
  We define a sub-dominant symmetry component as any attractive symmetry
channel which has a lower transition temperature than the bulk symmetry
channel.
  These effects were discussed in detail in [I] in connection with
the calculation of the self-consistent order parameter and free energy.
  It was shown there that the most prominent effect comes from the mixing
in of the $B_2$ symmetry component since it allows for an effective rotation
of the total order parameter towards its optimal (symmetric) configuration
at the wall.
  This has a very pronounced {\it qualitative} effect on the structure of
the order parameter near the wall, which is in turn discernible in the
density of states.
  In Fig.~6 we plot the superconducting DOS factor, $\nu$,
for a system consisting of a $B_1$ bulk phase coupled to a $B_2$ symmetry
component with a lower $T_c$.
  The choice of parameters is the same as in Fig.~4 for the
purpose of comparison.
  The zero energy bound states are still present in the $\phi=20\deg$ and
$\phi=45\deg$ curves since the formation of these states depends only on
the asymptotic values of $\Delta(\rho)$ which are not changed when surface
mixing is present (i.e.~the bulk phase is not changed).
  The pole in the $\phi=0\deg$ trajectory, however, has now vanished
since the rotation of the total order parameter gives rise to a local
maximum in the order parameter near the wall rather than a local minimum.
  On the other hand, an additional bound state now appears just below the
continuum in the $\phi=45\deg$ curve.
  The appearance of this pole is a result of the broadening of the local
minimum formed by the absolute value of the order parameter along the
$\phi=45\deg$ trajectory when mixing is allowed to take place.

  So far our discussion of the tunneling density of states has focused on
the value of the superconducting DOS factor, $\nu$,  at the wall, since this
is the case which is presumably accessible to tunneling experiments.
  For completeness, however, we show in Fig.~7 the behavior of $\nu$ as we
move into the sample.
  Here we have selected the $\phi=45\deg$ trajectory curve from Fig.~6
since it contains both bound state features.
  As we move away from the wall, $\nu$ approaches its bulk form by
continuously transferring weight from the bound states to the gap edge.
  The inset in Fig.~7 depicts the manner in which the bound state weights
decay as we move into the bulk.
  While the zero-energy bound state is highly localized near to the wall,
the upper bound state is spread out over a much broader region, which once
again emphasizes the distinct nature of these two states.
  The ripples that develop in the superconducting DOS factor for
$\epsilon>\Delta_{max-bulk}$ are Tomasch
oscillations\cite{tomasch65,tomasch66} whose prominence is a consequence of
the cleanliness of our system.

\begin{figure}
\centerline{\psfig{figure=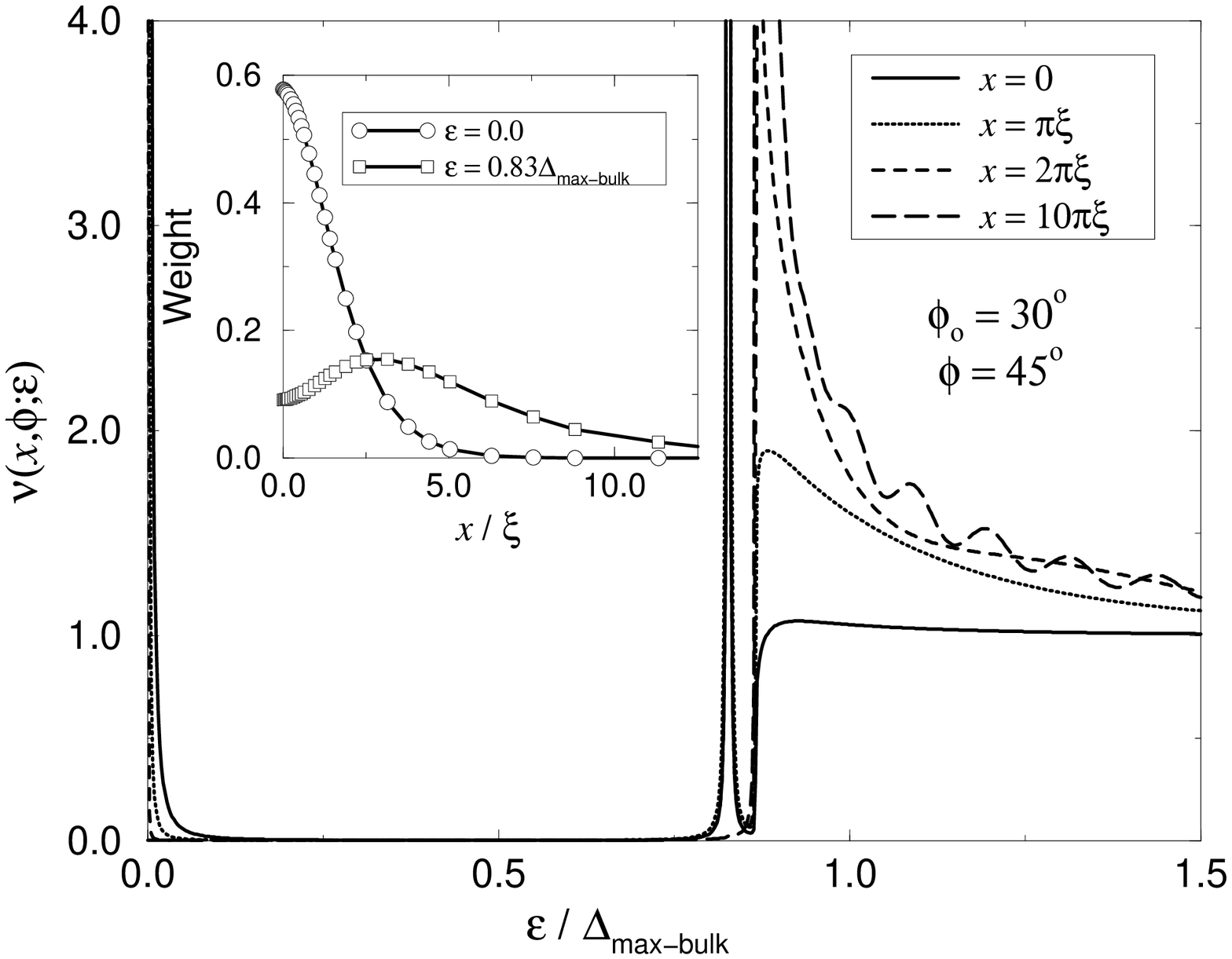,height=3.0in}}
\begin{quote}
\small
Fig.~7.~~
  The density of states for a $B_1$-symmetry order parameter coupled to
a sub-dominant $B_2$-symmetry component as a function of distance from
the wall.
  The inset shows the dependence on $x$ of the weight of the two bound
state delta functions (at $\epsilon=0.0$ and at $\epsilon=0.83$).
\end{quote}
\end{figure}

  The mixing in of an order parameter component transforming according to
the $A_2$ symmetry channel has also been considered.
  However, since the inclusion of an $A_2$ component only allows for a
{\it distortion} of the $B_1$ order parameter in momentum space (rather
than an overall {\it rotation}), we find very little variation in the
qualitative features of the density of states.

  We presented in this section our results for the angle-resolved
density of states near a surface.
  We close this section by discussing the I--V characteristic for quasiparticle
tunneling of an S--N junction where S refers to a $d$-wave superconductor.
  The differential tunneling conductance, $dI/dV$, is, at zero temperature,
given by the following equation\cite{duke69}:
\begin{equation}
  \frac{dI}{dV} = \frac{1}{R_N}\int_{-\pi/2}^{\pi/2} \, d\phi
                  D(\phi) \, \nu(\phi,eV),
\end{equation}
where $\nu(\phi,eV)$ is the normalized density of states
(see equation (4)), and $D(\phi)$ is the ``barrier function'' which is
normalized such that
\begin{equation}
  \int_{-\pi/2}^{\pi/2} d\phi \, D(\phi) = 1.
\end{equation}
  The function $D(\phi)$ describes the physical properties of the tunneling
barrier, and therefore depends on the details of the experimental
configuration.
  For example, a $D(\phi)=\delta(\phi)$ describes a ``thick'' barrier with
dominant tunneling into the forward direction.
  Whereas a $D(\phi)=1/\pi$ corresponds to a ``random'' barrier with
equal probabilities of tunneling into all directions.

\begin{figure}
\centerline{\psfig{figure=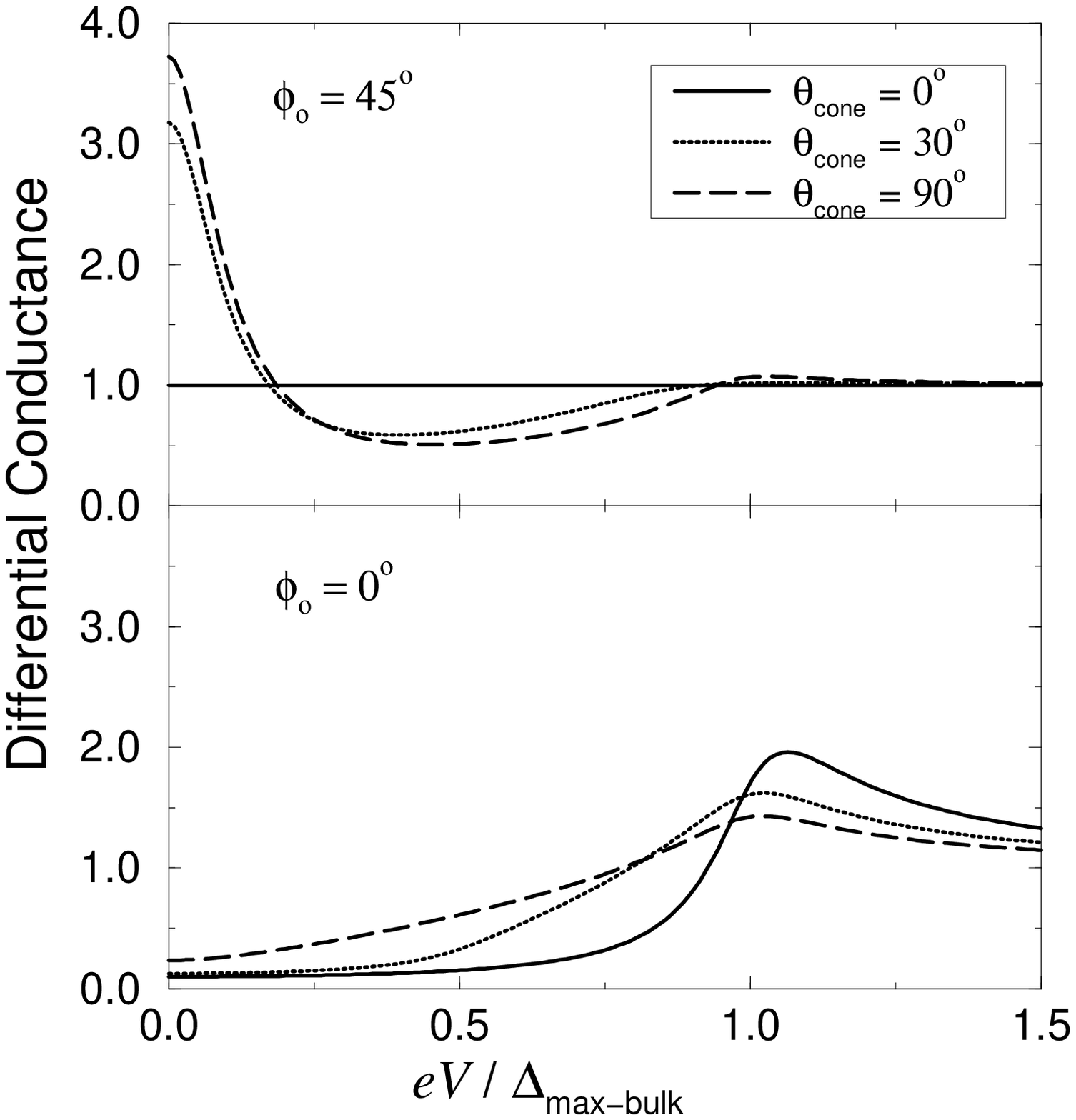,height=3.0in}}
\begin{quote}
\small
Fig.~8.~~
  The $T=0$ differential tunneling conductance as a function of bias
voltage, $eV$, for a (110) surface and a (100) surface and three different
sampling-cone sizes.
  An energy width ($\sim0.1\Delta_{max-bulk}$) has been included to simulate
experimental broadening.
\end{quote}
\end{figure}

  We present results for three types of model barriers.
  The barriers are described by a constant scattering probability into a
``sampling cone'' about the normal direction with an opening angle
$\theta_{cone}$.
  Specifically, we consider: (1) a thick barrier with $\theta_{cone}=0\deg$,
(2) a random barrier with $\theta_{cone}=90\deg$, and (3) an intermediate
barrier with $\theta_{cone}=30\deg$.
  In Fig.~8 we show the differential tunneling conductance for these three
barrier models considering the two most natural surface to lattice
orientations, i.e.~a [100] surface ($\phi_o=0\deg$) and a [110] surface
($\phi_o=45\deg$).
  Note that for $\phi_o=45\deg$ one finds a considerable zero-bias
anomaly\cite{hu94,kashiwaya95}, while for $\phi_o=0\deg$ one finds the
bulk density of states averaged over the sampling cone.
  Thus one expects, for $d$-wave superconductors with ideal surfaces, a
very different tunneling characteristic depending on surface orientation.
  We know of no experiment which has considered the tunneling spectrum as
a function of surface orientation and which would therefore be in a position
to either verify or refute these clear predictions of the $d$-wave model.

\begin{figure}
\centerline{\psfig{figure=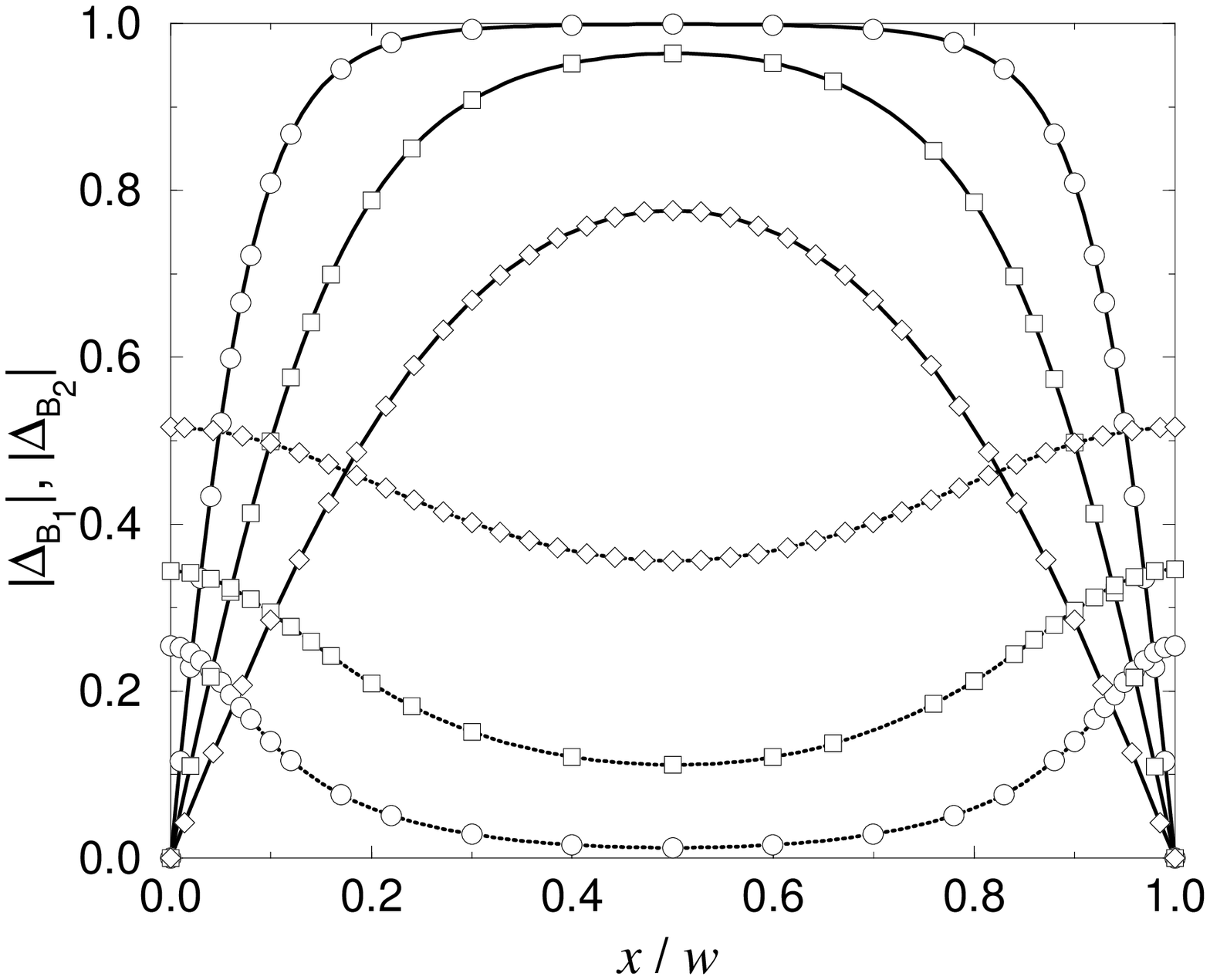,height=2.6in}}
\begin{quote}
\small
Fig.~9.~~
  Typical order parameter profiles in the slab geometry when both a $B_1$
and a $B_2$ component are present.
  The data are for the (1,1,0) surface ($\phi_o=45\deg$) of the slab.
  The solid (dotted) lines correspond to the amplitudes of the $B_1$
($B_2$) components.
  The curves are drawn for three different slab widths: $10\pi\xi$ (circles),
$5\pi\xi$ (squares), and $3.5\pi\xi$ (diamonds).
\end{quote}
\end{figure}

\subsection{Finite Slab}

  We consider a slab of finite thickness, $w$, and infinite lateral extent,
with two specular walls located at $x=0$ and $x=w$.
  This is a very convenient setting for studying surface effects since
a quasiparticle experiences the surface regions several times along its
trajectory.
  This enhances the effects of surfaces on the density of states.
  This system was studied in detail by Nagato {\it et.~al.}~\cite{nagato95}
for the case of a single-component $B_2$-symmetry order parameter, and a
surface to lattice tilt angle of $\phi_o=0\deg$.
  They discussed the existence of a phase transition from the
superconducting state to the normal state as a function of layer thickness.
  This scenario becomes more complicated, however, if we allow the mixing-in
of sub-dominant symmetry components.
  In this case we find, for certain tilt angles $\phi_o$, that a phase
transition involving only one of the symmetry components can occur, leaving
the system in a superconducting state of a different symmetry.
  In this way the normal state is not realized until one reaches much
smaller sample sizes, if at all.

\bigskip
\begin{figure}
\centerline{\psfig{figure=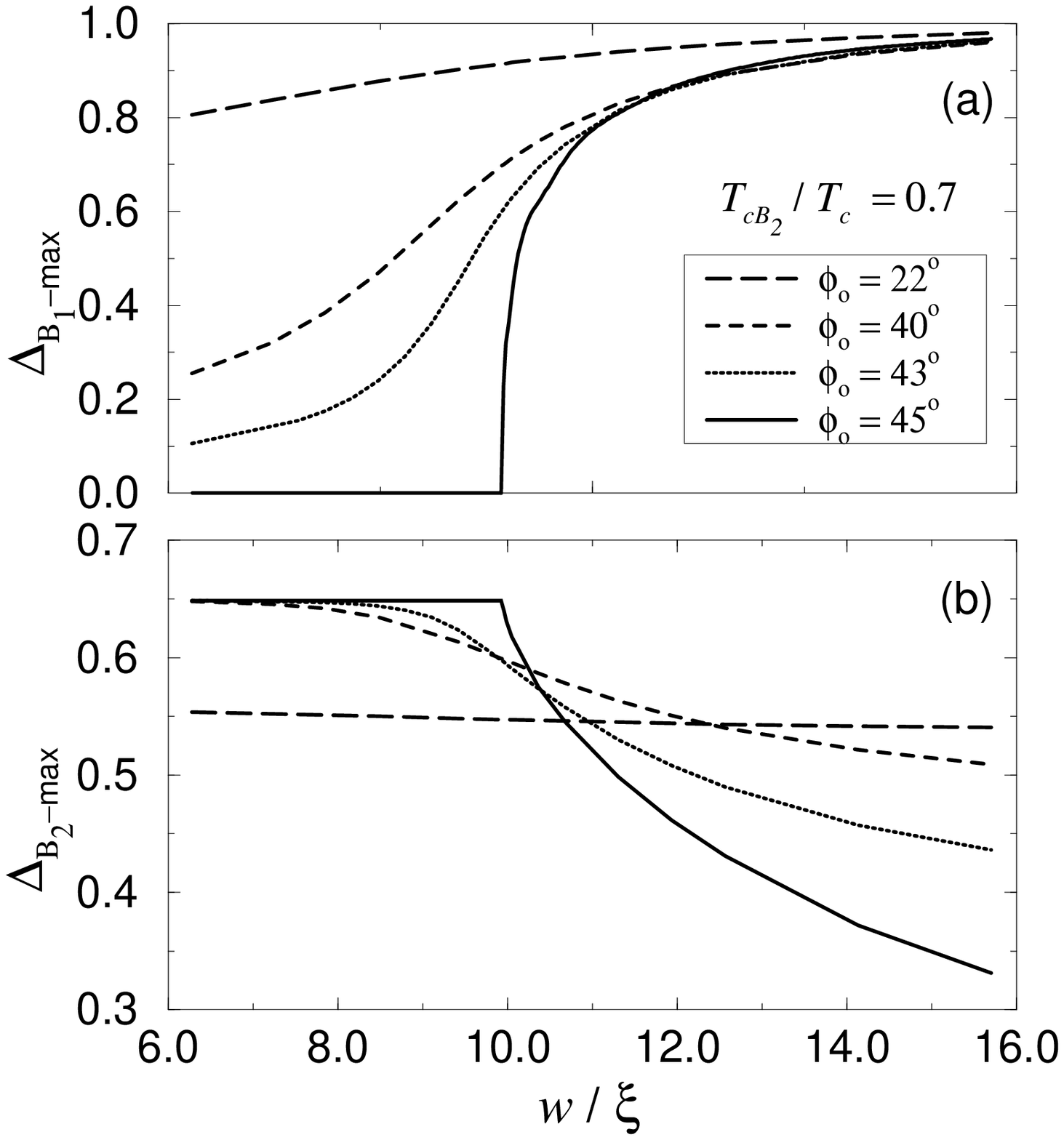,height=3.2in}}
\begin{quote}
\small
Fig.~10.~~
  The dependence of the order parameter components on the width of the
slab: (a) The maximum of the $B_1$-symmetry component for various slab
to lattice tilt angles, $\phi_o$, and (b) the maximum of the $B_2$-symmetry
component for the same tilt angles.
  Note that the system has a phase transition at a slab width $w=9.9\xi$
for $\phi_o=45\deg$.
\end{quote}
\end{figure}

  In order to exhibit these effects we consider the case of an isotropic
cylindrical Fermi surface and compute the self-consistent order parameter
in the same manner as discussed in [I].
  We again assume that the $B_1$ symmetry channel possesses the highest
transition temperature, and allow for the mixing in of a $B_2$ symmetry
component with $T_{cB_2}/T_c = 0.7$.
  In Fig.~9 we plot a typical set of order parameter profiles for a
surface to lattice orientation of $\phi_o=45\deg$ and various slab widths.
  As the sample becomes thinner the $B_1$ component is strongly suppressed,
while the presence of the $B_2$ component becomes enhanced.
  A plot of the maximum absolute value of each of the components as a
function of the slab width is shown in Fig.~10 for several surface to
lattice orientations.
  The angle $\phi_o=45\deg$ is a special case since it represents the maximal
pair-breaking orientation for the $B_1$ component and the minimal pair-breaking
(i.e.~symmetric) orientation for the $B_2$ component.
  Hence, the $B_1$ component is favored in the bulk since it possesses the
highest $T_c$, however the $B_2$ component is favored at the wall for symmetry
reasons.
  As a result, at a width $w\approx9.9\xi$ we find a second order phase
transition in which the $B_1$ component abruptly vanishes leaving behind a
spatially uniform $B_2$-symmetry order parameter.
  As we move to angles away from $\phi_o=45\deg$, we find that the phase
transition is smeared out.

  In Fig.~11 we plot the superconducting DOS factor, $\nu$, in the vicinity
of the phase transition for two different quasiparticle trajectories.
  The dotted lines show the form of $\nu$ below the critical width and are
just what one expects for a spatially uniform order parameter (i.e.~they are
identical to the bulk results).
  Above the critical width the spectrum exhibits a complicated band-like
structure\cite{nagato95}.
  Since we are considering a clean system with specular walls, a quasiparticle
will propagate without dissipation, reflecting an infinite number of times,
always with the same angle of incidence.
  The local order parameter along such a trajectory has a periodic structure
so that this scenario closely parallels the case of a particle in a periodic
potential.
  The splitting of the excitation spectrum into bands is thus a consequence
of Bloch's theorem.

\bigskip
\begin{figure}
\centerline{\psfig{figure=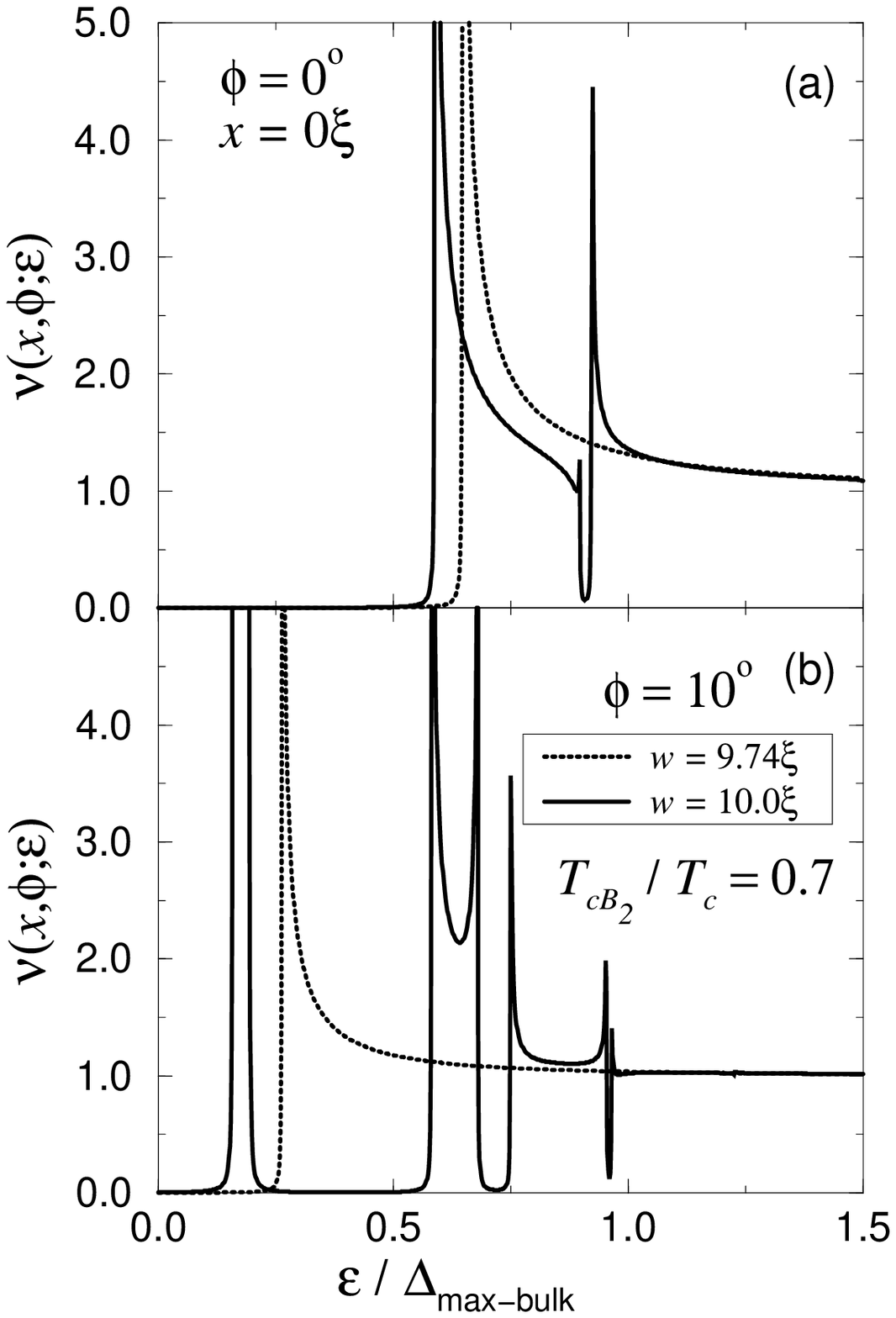,height=3.6in}}
\begin{quote}
\small
Fig.~11.~~
  The superconducting density of states in a slab with a slab to lattice
orientation $\phi_o=45\deg$.  Results are shown slightly above ($w=10.0\xi$)
and below ($w=9.7\xi$) the phase transition for two different trajectories:
(a) $\phi=0\deg$, and (b) $\phi=10\deg$.
\end{quote}
\end{figure}

  We have also considered the case of a $B_1$-symmetry order parameter
coupled to a sub-dominant $A_2$-symmetry component.
  In this case it is the tilt angle $\phi_o=22.5\deg$ which is special,
since it represents the symmetric orientation for the $A_2$ component.
  Reminiscent of the previous case, we again observe a phase transition
around $w\sim5.5\xi$ in which the $B_1$ component vanishes while the $A_2$
component becomes spatially uniform.
  However, here the transition is first order and, moreover, several
distinct near-by metastable states compete.
  In the transition region, the differences in free energy between the
various states are so small that specific perturbations in any experimental
configuration would surely be decisive.
  The superconducting DOS factor, $\nu$, exhibits distinguishing features
on either side of the transition which, again, just parallels the previous
case.

\section{Conclusions}

  We have studied the spectral properties of an anisotropically paired
superconductor in the vicinity of a superconductor-insulator interface
by considering two different physical geometries: a semi-infinite system,
and a slab of finite width and infinite lateral extent.
  The stable bulk phase of the order parameter is always taken to transform
like the $B_1$ ($d_{x^2-y^2}$) representation of the $D_{4h}$ (tetragonal)
group, however other attractive symmetry channels are also allowed to be
present, which has a pronounced effect on the order parameter's behavior
near the walls.
  The calculation of the self-consistent order parameter for systems such
as these was discussed by the current authors in
Ref.~\onlinecite{buchholtz95} and, we have carried over those results as
input into the current calculations.
  All of our calculations were performed considering both an isotropic
cylindrical Fermi surface, as well as a more realistic tight-binding Fermi
surface; we find very little difference in the qualitative features of our
results between the two Fermi surface models.

  The spectra at the surface of a semi-infinite system exhibit three
characteristic features.
  First, one finds band-edge singularities at energies corresponding to
the bulk gap magnitudes $|\Delta_{bulk}(\p_{f,\mbox{in}})|$ and
$|\Delta_{bulk}(\p_{f,\mbox{out}})|$ which yield information about the gap
anisotropy far from the wall.
  Secondly, one observes the formation of surface bound states which are
of two distinct types.
  For any order parameter which changes its sign along the Fermi surface
(regardless of its symmetry), one can always find situations in which a
bound state will be formed at zero energy\cite{hu94,atiyah75}.
  The formation of these states depends only on the relative sign of
the asymptotic order parameter values ($\Delta_{bulk}(\p_{f,\mbox{in}})$ and
$\Delta_{bulk}(\p_{f,\mbox{out}})$) along a quasiparticle trajectory.
  As a consequence, the mixing-in of sub-dominant symmetry components with
the bulk ($B_1$) symmetry component has no impact on the presence of these
states, since the order parameter is only altered within a few coherence
lengths of the surface.
  Depending on the detailed nature of the interaction of the order
parameter with the wall, however, a second species of bound states may
appear at higher energies, if the order parameter is sufficiently
suppressed in the surface region.
  The weight and location (in energy) of these bound states is directly
related to the spatial dependence of the order parameter near the wall.
  Thus, one finds that the formation of these states can be correlated to
both the strength and symmetry of the sub-dominant components.
  All of these features depend strongly on the surface to lattice
orientation, and thus one can conceive of mapping out the full anisotropy
of the order parameter by measuring the surface density of states for
various surface orientations.

  The study of a $d$-wave order parameter in a slab geometry can be
especially illuminating since the percentage of the sample volume which lies
in the surface region is significantly enhanced.
  For the case of single $d$-wave order parameter in a finite slab, one
finds phase transitions for certain surface to lattice orientations which
results in a nontrivial temperature vs. width phase diagram\cite{nagato95}.
  When the bulk $B_1$ symmetry order parameter is allowed to couple to a
$B_2$ symmetry component, these phase transitions are washed out for most
surface orientations.
  For certain ``special'' surface orientations however, phase transitions
still occur (of either first or second order) which may take the sample into
the normal state or even into a state with a different bulk symmetry.
  These special orientations are usually those in which the sub-dominant
component does {\it not} pair-break.
  These phase transitions have a pronounced effect on the excitation spectrum
since the order parameter always has a complicated spatial structure on one
side of the transition while it is completely structureless on the other.

  The calculations presented in this paper were carried out for clean
superconductors with perfectly reflecting surfaces.
  This is important to note because the presence of a diffuse boundary
could result in significant qualitative effects.
  Indeed, our ongoing work indicates that for diffuse surfaces the
distinction between the various surface to lattice orientations may no
longer be as clear, since any one in-coming trajectory may mix with all
out-going trajectories.
  This effect tends to smear-out, to variable degrees, some of the
aforementioned phase transitions and characteristic features in the
surface excitation spectrum.
  We conclude that a concerted effort should be made to fabricate ``clean''
surfaces, with specific surface to lattice orientations, if the maximal
quantity of information on the anisotropy of the order parameter and the
surface induced features in the density of states is to be extracted.

\section*{Acknowledgements}

  The research of L.J.B was supported by the Fulbright Commission and that
of M.P. by the Alexander von Humboldt-Stiftung.
  D.R. was supported, in part, by the Graduiertenkolleg ``Materialien und
Ph\"anomene bei sehr tiefen Temperaturen'' of the DFG.
  J.A.S. acknowledges partial support by the Science and Technology Center
for Superconductivity through NSF Grant no.~91-20000.
  Authors D.R. and J.A.S. also acknowledge additional support from the
Max-Planck-Gesellschaft and the Alexander von Humboldt-Stiftung.

\end{document}